\documentclass[12pt]{iopart}

\usepackage{hyperref}
\usepackage{graphicx}
\usepackage{iopams}
\usepackage[numbers]{natbib}

\begin{document}

\title[Probing a theoretical framework for a Photonic Extreme Learning Machine]{Probing a theoretical framework for a Photonic Extreme Learning Machine}

\author{Vicente Rocha$^{1,2}$, Duarte Silva$^{3}$, Felipe C. Moreira$^{1}$, Catarina S. Monteiro$^{1}$, Tiago D. Ferreira$^{1}$, Nuno A. Silva$^{1}$}

\address{$^1$ INESC TEC, Centre of Applied Photonics, Rua do Campo Alegre 687, 4169-007 Porto, Portugal}
\address{$^2$ Faculty of Sciences of the University of Porto, Department of Physics and Astronomy, Rua do Campo Alegre 687, 4169-007 Porto, Portugal}
\address{$^3$ Department of Electrical Engineering, Eindhoven University of Technology, 5600 MB Eindhoven, the Netherlands} 

\ead{nuno.a.silva@inesctec.pt}
\vspace{10pt}
\begin{indented}
\item[]August 2024
\end{indented}

\begin{abstract}

The development of computing paradigms alternative to von Neumann architectures has recently fueled significant progress in novel all-optical processing solutions. In this work, we investigate how the coherence properties can be exploited for computing by expanding information onto a higher-dimensional space in the photonic extreme learning machine framework. A theoretical framework is provided based on the transmission matrix formalism, mapping the input plane onto the output camera plane, resulting in the establishment of the connection with complex extreme learning machines and derivation of upper bounds for the hidden space dimensionality as well as the form of the activation functions. Experiments using free-space propagation through a diffusive medium, performed in low-dimensional input space regimes, validate the model and the proposed estimator for the dimensionality. Overall, the framework presented and the findings enclosed have the potential to foster further research in a multitude of directions, from the development of robust general-purpose all-optical hardware to a full-stack integration with optical sensing devices toward edge computing solutions.


\end{abstract}


%

\section{Introduction}

Optical computing has long been promising applications for faster information processing and lower power consumption~\cite{athale2016optical}. Today, photonic hardware contributes to modern systems in multiple tasks, from accelerating data transmission~\cite{richardson2013space} and implementing specific linear operations such as Fourier transforms~\cite{goodman2005introduction}, to more complex optical correlators~\cite{lugt1964signal}, integral transforms~\cite{athale1982optical}, matrix-vector multiplication~\cite{shen2017deep,spall2020fully}, and reconfigurable linear operators~\cite{matthes2019optical, chang2018hybrid, wu2021programmable}. Despite these successes, scalable and energy-efficient optical transistors and memories to support von Neumann general-purpose photonic processors remain ellusive due to weak light interaction and optical field confinement~\cite{kazanskiy2022optical}. One way to bypass these obstacles is to seek alternative computing frameworks, such as neuromorphic-inspired architectures, a research line that recently seeded the development of computing paradigms closer to analog computing, where light has long been known to excel.

In short, neuromorphic architectures offer a promising path for optical processors by leveraging a distributed memory framework through weights~\cite{hopfield1982neural, farhat1985optical} and approximation capabilities of neural networks (NN) enabled by the expressivity of non-linear maps~\cite{hornik1991approximation}. Together with native light speed and broad bandwidth of photonic hardware, it thus enables massive parallelism and throughput in compact, low-power devices~\cite {wetzstein2020inference}. Recent advances in photonic neuromorphic computing have followed several promising directions, with one of the most compelling paths being the realisation of all-optical deep neural networks through a sequential series of fabricated~\cite{lin2018all, luo2022metasurface} or reprogrammable diffractive layers~\cite{liu2022programmable, zheng2022optimize}. Although fabricated layers have the shortfall of fixing the setup for a specific task, using reprogrammable layers unlocks hardware change between tasks, giving rise to a versatile family of task-specific accelerators that take advantage of physics-aware training and even structural nonlinearities~\cite{abou2025programmable}. However, as the size scale of the network scales, accumulated optical loss, noise, fabrication imperfections, and the time and energy overhead of training these devices result in limited practical scalability.

A greater degree of flexibility is offered by reservoir computing (RC) and, in particular, its memoryless counterpart, the Extreme Learning Machine (ELM)~\cite{huang2004extreme, Huang2006ExtremeLM,li2005fully}. In these architectures, the combination of random weights and the nonlinear activation function gives the framework a universal function approximation capability for both regression and classification problems~\cite{Huang2006ExtremeLM}. The photonic implementation of ELMs, coined Photonic Extreme Learning Machine (PELM), has been demonstrated in free-space architectures~\cite{pierangeli2021photonic, m2022large}, optical fibers~\cite{redding2024fiber}, and integrated optics~\cite{biasi2023array}, typically exploiting the inherent randomness of light scattering to realise a fixed random weight matrix. Programmable optical hardware has also been employed, enabling ensemble techniques and evolutionary optimisation strategies~\cite{rausell2025programmable}. These systems have already tackled tasks ranging from dimensionality reduction~\cite{liutkus2014imaging} and kernel approximation~\cite{saade2016random} to natural language processing~\cite{m2022large}. However, the overall comprehension of the ultimate limits and opportunities of these optical computing devices would benefit from a more comprehensive theoretical study on the dimensionality of the output space and its activation functions, in particular when the input space has low dimensionality.

\begin{figure}
\makebox[\textwidth]{
\includegraphics[width=1.\linewidth]{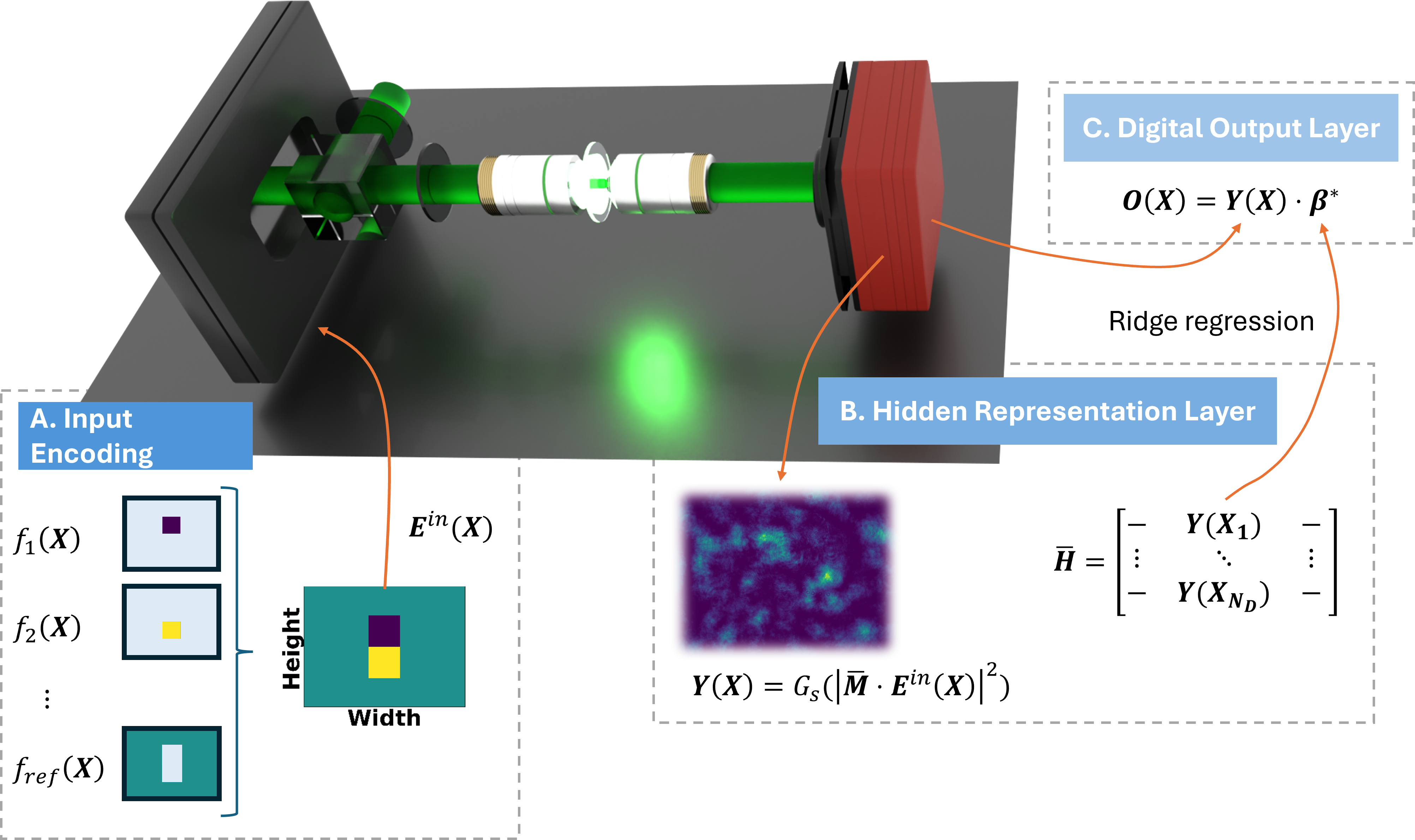}
}
\caption{Overview of the optoelectronic PELM experimental implementation and connection to the theoretical framework. \textbf{A.} The information of each input state $\boldsymbol{X}$ is encoded on the incident optical beam in phase or amplitude using an SLM. \textbf{B.} The beam is then focused on an optical media (here a diffusive media) inducing a random transmission matrix $\boldsymbol{\bar{M}}$ modeling the propagation before being collimated again resulting in a speckle patter then collected with a CMOS sensor, resulting in the hidden layer representation $\boldsymbol{Y}\left(\boldsymbol{X}\right)$. \textbf{C.} The imaged speckle pattern is then used as input to the linear model $\boldsymbol{O}(\boldsymbol{X}) = \boldsymbol{Y}\left(\boldsymbol{X}\right) \boldsymbol{\bar{\beta}}^*$, obtaining a prediction $\boldsymbol{O}\left( \boldsymbol{X}\right)$ by leveraging on a $\boldsymbol{\beta}^*$ trained with Ridge regression using the hidden layer output matrix $\boldsymbol{\bar{H}}$ built from $N_D$ training samples.}
\label{fig:OIComputer}
\end{figure}

Taking this context as the starting point, this work analyzes a simplified free-space optical computing configuration (depicted in Figure~\ref{fig:OIComputer}) that serves as an experimental PELM test bed. First, in section~\ref{sec: theoretical}, we develop a theoretical model by describing both the input information encoding, the random linear propagation using the transmission matrix formalism, and the detection step, finishing with the derivation of the dimensionality upper bounds of the hidden space for linear camera response. Then, we proceed by proposing a methodology to estimate the dimensionality of the space in the presence of experimental noise, via the analysis of the spectrum of singular values. In section~\ref{sec: experimental} we employ the experimental setup in two low-dimensional spaces to validate the theoretical model and observe the enrichment of the hidden space by the introduction of the camera nonlinear response. To conclude, in section~\ref{sec: discussion} we discuss future prospect of growth for the expressivity of this technology based on the theoretical analysis provided.
    
\section{\label{sec: theoretical} Theoretical Framework}

The optical computing setup used for this work depicted in Figure~\ref {fig:OIComputer} permits a simplified, stage-by-stage analysis of encoding, transmission, and detection within a typical PELM implementation. Briefly, the input features of a dataset are encoded on the wavefront of a coherent optical beam using a spatial light modulator (SLM) either with phase or amplitude modulation, as described in the next subsections. As the information in the wavefront propagates, it is randomly mixed by the tight focusing of a microscope objective onto a diffusive medium, resulting in a speckle pattern. A second objective collimates the light that is then collected by a CMOS camera. The integrated intensity across $N_{out}$ macropixel regions of the sensor composes the hidden layer used to train a final digital linear transformation to solve a task (regression or classification problem).
    
In the discussion that follows, we model how information is transformed as the optical signal propagates through each stage. This theoretical model aims to clarify how the interplay of input encoding, scattering-induced randomness, and measurement collectively determines the activation functions and effective dimensionality of the hidden layer, imposing limitations on its computing capacity for complex computations.


\subsection{Input Layer and feature encoding}

In the proposed system, an input vector $\boldsymbol{X} \in \mathbb{R}^{N_I}$ containing $N_I$ independent and non-constant features is encoded on the wavefront by modulating the amplitude and phase of $N_{enc}$ spatial modes $\{\boldsymbol{e}^{(in)}_j, j=1,\cdots, N_{enc} \}$ (see Figure ~\ref {fig:OIComputer} for details), plus an additional tunable reference mode. Each spatial channel is assigned a complex encoding function $f_j(\boldsymbol{X}) = a_j(\boldsymbol{X}) \exp\left[i b_j(\boldsymbol{X})\right]$ with amplitude $a_j(\boldsymbol{X}) \in [0, 1]$ and $b_j(\boldsymbol{X}) \in [0, 2\pi]$. With this encoding, we must highlight that PELM performs the computation in the codomain of $a\left(\boldsymbol{X}\right) \exp\left[i b_j(\boldsymbol{X})\right]$ rather than $\boldsymbol{X}$ itself.
    
The information in $\boldsymbol{X}$ resides now in a continuous optical field, but the dimensionality of this space remains to be determined because it depends on the specific choice of the encoding functions and their spatial distribution. Collecting the spatial mode basis $\boldsymbol{\bar{e}}^{\left(in\right)} = \left[\boldsymbol{e}_1^{\left(in\right)} \dots \boldsymbol{e}_{N_{enc}}^{\left(in\right)}\right]$ and the corresponding encoding vector $\boldsymbol{F}=\left[f_1\left(\boldsymbol{X}\right),\dots, f_{N_{enc}}\left(\boldsymbol{X}\right)\right]^T$, allows us to express the input field as $\boldsymbol{E}^{(in)}(\boldsymbol{X}) = \boldsymbol{\bar{e}}^{\left(in\right)} \boldsymbol{F} \left(\boldsymbol{X}\right)$. To simplify implementation and avoid pre‑processing that mixes feature information, we adopt a one‑to‑one mapping between input dimensions and spatial modes, i.e. $f_i$ depends only on the $i$‑th entry of $\boldsymbol{X}$. With this separable encoding, the number of encoded modes is $N_{enc} = N_I + 1$.

\subsection{Linear Propagation in Scattering Media}

The propagation of the optical beam in the scattering media is well described using the transmission matrix formalism~\cite{popoff2010measuring} establishing a linear mapping between the encoding plane and the detection plane. Discritising the input and output apertures into $N_{enc}$ and $N_{out}$ channels (number of pixels or macro-pixels used for detection) we may define a complex transmission matrix $\boldsymbol{\bar{M}} \in \mathbb{C}^{N_{out} \times N_{enc}}$ that links each input mode to every output mode. The optical field reaching the camera plane is therefore $\boldsymbol{E}^{\left(out\right)} = \bar{\boldsymbol{M}} \cdot \boldsymbol{F}\left(\boldsymbol{X}\right)$, a linear combination of the encoding functions.

By discretising the input and output plane using the transmission matrix formalism, we cast the optical mapping in linear algebra terms. For diffusive media, the matrix $\boldsymbol{\bar{M}}$ obeys random-matrix statistics~\cite{popoff2010measuring} and thus can be assumed full-rank, i.e. $rank\left(\boldsymbol{\bar{M}}\right)=\min\left(N_{out}, N_{enc}\right)$. This leads to two distinct regimes. The first, dimensionality reduction, arises when $N_{enc} > N_{out}$. Here, each output feature is independent of the others, but the rank of the transmission matrix induces a bottleneck in the dimensionality of the field. Conversely, when $N_{enc}\leq N_{out}$, the scattering stage mixes the input features while preserving the dimensionality of the field and is capable of supporting a higher-dimensional space. For this reason in this manuscript, we focus on the second scenario.

\subsection{Hidden Layer Output}

The next stage of the implemented optical PELM is the detection at the camera plane, which will be the source of nonlinearity of the system. Indeed, each pixel will integrate contributions from the entire input aperture combined via the transmission matrix, meaning that a square-law detector records an interferometric speckle pattern of the input field. Formally, the optical read-out intensity is obtained by applying a composite function $G$ to the output field in an element-wise manner,
\begin{equation}
    \boldsymbol{Y}\left(\boldsymbol{X}\right) = G\left( \boldsymbol{E}^{\left(out\right)} \right) = G_s\left( I \left( \boldsymbol{E}^{\left(out\right)} \right) \right) ,
\label{eq: output intensity}
\end{equation}
where the interference of coherent waves $I(\boldsymbol{E}^{\left(out\right)}) = |\boldsymbol{E}^{\left(out\right)}|^2$ results in an optical speckle that is separated from the camera optoelectronic response $G_s$, which can potentially introduce further non-linearity.

For a direct comparison with the formalism of ELMs, we reinterpret equation~\ref{eq: output intensity} by partitioning the feature vector into a data-dependent part and a constant reference, $\boldsymbol{F}(\boldsymbol{X}) = \boldsymbol{F}^{'}(\boldsymbol{X})  + \boldsymbol{C}$, where $\boldsymbol{F}^{'} = \left[ f_1 (x_1), \dots, f_{N_{input}}(x_{N_{input}}), 0 \right] ^T$ and $\boldsymbol{C} = \left[0,\dots,0, f_{ref} \right] ^T$. Substituting into the optical read‑out gives $\boldsymbol{Y}\left(\boldsymbol{X}\right) = G\left(\boldsymbol{\bar{M}}\boldsymbol{F}'\left(\boldsymbol{X}\right) + \boldsymbol{b}\right)$ which is has the conventional form of neural networks by explicitly separating the linear mapping of the input features from the bias $\boldsymbol{b} = \boldsymbol{\bar{M}}\boldsymbol{C}$ vector contributed by the tunable reference mode and randomised by the transmission matrix.

In the study of ELMs, the central object for analysis is the hidden layer output matrix, which we build from the optical read-out as
\begin{equation}
    \boldsymbol{\bar{H}} = \left[ \begin{array}{ccc}
         \rule[.5ex]{2.5ex}{0.5pt} & G\left( \boldsymbol{\bar{M}} \boldsymbol{F} '\left(\boldsymbol{X}_{1}\right) + \boldsymbol{\bar{M}} \boldsymbol{C}\right) & \rule[.5ex]{2.5ex}{0.5pt}\\
         & \vdots & \\
         \rule[.5ex]{2.5ex}{0.5pt} & G\left( \boldsymbol{\bar{M}} \boldsymbol{F} '\left(\boldsymbol{X}_{N_D}\right) + \boldsymbol{\bar{M}} \boldsymbol{C}\right) & \rule[.5ex]{2.5ex}{0.5pt}
\end{array} \right],
\label{eq: photonic H}
\end{equation}
with $N_{out}$ columns and $N_D$ rows. In short, from ELM theory (see~\ref{ELM framework} and details in reference~\cite{Huang2006ExtremeLM}) provided that:
\begin{itemize}
    \item the activation function $G$ in $\boldsymbol{\bar{H}}$ is a nonlinear, non-polynomial function that is infinitely differentiable over the input domain; and
    \item the weights drawn from a continuous distribution,
\end{itemize} 
then the matrix $\boldsymbol{\bar{H}}$ has full rank and the ELM satisfies the universal approximation property (i.e. may arbitrarily approximate a given continuous function to a desired level of accuracy).

\subsection{Digital Output Layer}

The final stage of the architecture is the output prediction $\boldsymbol{O}(\boldsymbol{X}) \in \mathbb{R}^{N_y}$ obtained from the hidden-layer read-out via the linear model $\boldsymbol{O}\left(\boldsymbol{X}\right) = \boldsymbol{Y}\left(\boldsymbol{X}\right) \boldsymbol{\bar{\beta}}$, where $\boldsymbol{\bar{\beta}} \in \mathbb{R}^{N_{out}\times N_y}$ is the output weight matrix and with $N_y$ being the dimensionality of the target $\boldsymbol{T}(\boldsymbol{X}) \in \mathbb{R}^{N_y}$ associated to each input vector. In the formal ELM framework, the weights are learned in a supervised manner by minimising the $\ell_2$ norm featuring a unique global minimum obtained from inversion of the hidden layer output matrix. To preserve all native optical computing advantages, the digital readout involves only linear operations. Thus, it does not introduce further nonlinear transformations, and the expressive power of the overall model remains rooted in the optical stages described above.

To formalise the supervised training step for obtaining the weight matrix, we gather $N_D$ labels into a target matrix, ordered row-wise as
\begin{equation*}
    \boldsymbol{\bar{T}} = \left[ \begin{array}{ccc}
        \rule[.5ex]{2.5ex}{0.5pt} & \boldsymbol{T}\left(\boldsymbol{X}_1\right) & \rule[.5ex]{2.5ex}{0.5pt} \\
        & \vdots & \\
        \rule[.5ex]{2.5ex}{0.5pt} & \boldsymbol{T}\left(\boldsymbol{X}_{N_D}\right) & \rule[.5ex]{2.5ex}{0.5pt}
\end{array} \right],
\label{eq: dataset}
\end{equation*}
so that each sample corresponds to the respective hidden layer row in $\boldsymbol{\bar{H}}$. For regression tasks, we can minimise a regularised empirical metric that balances the fit of the model with the norm of the weight matrix. Given the hidden-layer output matrix $\boldsymbol{\bar{H}}$ and the target matrix $\boldsymbol{\bar{T}}$, the optimal weights are obtained from
\begin{equation*}
    \boldsymbol{\bar{\beta}}^* =  \mathop{\mathrm{argmin}}_{\boldsymbol{\bar{\beta}}}\left[\left\Vert \boldsymbol{\bar{H}} \boldsymbol{\bar{\beta}} -\boldsymbol{\bar{T}}\right\Vert _{q}^{\alpha_{2}} + \lambda\left\Vert \boldsymbol{\bar{\beta}}\right\Vert _{p}^{\alpha_{1}} \right],
\end{equation*}
where $\alpha_{1},\alpha_{2}>0$, $p,q=0,1/2,1,2,\ldots,+\infty$ specify the matrix norms, while $\lambda>0$ controls the strength of the regularisation. 
Among the available regularisation schemes, Ridge regression with $\alpha_{1}=\alpha_{2}=p=q=2$ is particularly attractive because it admits a closed-form solution~\cite{hastie2009elements}
\begin{equation}
\boldsymbol{\bar{\beta}}^* = \left( \boldsymbol{\bar{H}}^T \boldsymbol{\bar{H}} + \lambda \boldsymbol{\bar{1}}\right)^{-1} \boldsymbol{\bar{H}}^T \boldsymbol{\bar{T}},
\label{eq: ridge solution}
\end{equation}
and the Ridge parameter $\lambda$ improves generalisation by penalising the $\ell_2$ norm of the weights~\cite{bartlett2002sample}, being $\boldsymbol{\bar{1}}$ the identity matrix. This analytic expression yields the unique global minimiser in a single matrix inversion, avoiding the convergence issues of iterative solvers. 
        
In classification settings, one can repurpose Ridge regression by treating the output $\boldsymbol{O}\left(\boldsymbol{X}\right)$ as a vector of class scores. The predicted label can then be obtained as the largest component, allowing the model to serve as a multi-class classifier without additional post-processing. For binary tasks, a single output dimension suffices, with targets encoded as $-1$ and $1$, and the sign determines the predicted class.

\section{Dimensionality of the hidden space and Universal approximation capability of the PELM}

As established before, the rank of the hidden-layer matrix $\boldsymbol{\bar{H}}$ determines the dimensionality of the hidden feature space having consequences on the computing capabilities of ELMs. To correctly estimate the dimensionality with the rank of $\boldsymbol{\bar{H}}$, the matrix must capture the full range of degrees of freedom in this space. A practical bottleneck arises when the training set undersamples the hidden layer, $N_D < N_{out}$. Therefore, for now, we assume the minimal condition $N_D \geq N_{out}$, ensuring each output dimension is supported by at least one observation.
    
To expose the combinatorial nature of the hidden space, we expand equation \ref{eq: photonic H} by inserting the interferometric interaction and pulling the camera response $G_s$ outside the matrix as its action is element‑wise. Expressing the transmission matrix $\boldsymbol{\bar{M}}$ and the encoding matrix $\boldsymbol{\bar{F}}$ entry-wise reveals the row‑wise repetition of transmission coefficients and a column‑wise repetition of products of encoding functions. After straightforward algebra, it can be shown that the interferometric process is given by the sum of rank-1 matrices with the form
\begin{equation*}
\hspace{-1.5cm}
    \boldsymbol{\bar{H}}  = G_{s}\left(
        \sum_{j=1}^{N_{enc}} \sum_{k=1}^{N_{enc}} \left[\begin{array}{c}
            f_i\left(\boldsymbol{X}_1\right)  f^*_j\left(\boldsymbol{X}_1\right)  \\
            \vdots \\
            f_i\left(\boldsymbol{X}_{N_D}\right)  f^*_j\left(\boldsymbol{X}_{N_D}\right)
    \end{array} \right]
    \cdot
    \left[ m_{1,j}m^{*}_{1,k}, \dots, m_{N_{out},j}m^{*}_{N_{out},k}\right]
\right),
\end{equation*}
where $m_{i,j}$ denotes the $j$-th element of row $i$ of $\boldsymbol{\bar{M}}$. Each element of the interferometric hidden layer is therefore a randomly weighted sum of all pairwise products of the $N_{enc}$ encoded modes.

In the special case of a linear camera response ($G_{s}(x)=x$), the expression above simplifies to a sum of rank‑1 matrices in which the encoding functions are fully decoupled from the random transmission matrix. Assuming the maximum rank of the transmission matrix, the hidden layer matrix is then limited by the number of independent combinations of encoding functions and their conjugates, leading to the maximal bound $rank\left(\boldsymbol{\bar{H}}\right) \leq \min \left(N_{enc}^2, N_{out}\right)$ when both amplitude and phase encoding is used. Analogously to the output‑field analysis, a bottleneck can still arise from an insufficient number of output channels.

These thresholds can be derived analytically for the specific schemes of amplitude-only or phase-only modulation, resulting in smaller upper bounds for the hidden space dimensionality. In the case of amplitude-only modulation, the encoded functions are real-valued, therefore satisfying the identity $f_i\left(\boldsymbol{X}\right) = f_i^*\left(\boldsymbol{X}\right)$. Pairwise products satisfy $f_i f_j^* = f_i^* f_j$, leading to redundancy. The number of linearly independent terms equals the unordered pairs of encoding functions (including self‑products), giving the rank bound as the number of combinations
\begin{equation}
    rank\left(\boldsymbol{\bar{H}}\right) = C^{N_{enc}+1}_2.
\label{eq: amp rank}
\end{equation}
Besides, it is easy to conclude that for a linear response detection, amplitude modulation will lead to quadratic-polynomial activation functions.

In the case of phase-only modulation, the encoded functions are complex-valued so that $f_i\left(\boldsymbol{X}\right) \neq f_i^*\left(\boldsymbol{X}\right)$. Self-products reduce to a constant, $f_i\left(\boldsymbol{X}\right) f_i^*\left(\boldsymbol{X}\right) = constant$, whereas cross-products $f_i f_j^*$ for $i\neq j$ and their conjugates $f_j f_i^*$ are distinct functions. The maximum rank is therefore
\begin{equation}
    rank\left(\boldsymbol{\bar{H}}\right) = 2C^{N_{enc}}_2 +1.
\label{eq: phs rank}
\end{equation}
representing all distinct cross‑combinations plus one constant term. Following straightforward calculations, it is also possible to demonstrate that phase-only modulation will lead to cosine or sine-type activation functions, with a fixed frequency defined in the input space.

It is then possible to conclude that the linear camera model leads to an analytic upper bound on the hidden layer dimensionality. Under either modulation scheme, the hidden activations reduce to polynomials of the encoding functions. As a result, $\boldsymbol{\bar{H}}$ may be rank deficient if the number of input features is not sufficiently high (e.g. for low-dimensional input spaces) and thus the resulting PELM fails to satisfy the requirements set forth by ELM universal approximation theorem. 

Departing from the linear detector response, it is easy to conclude also that a saturated camera regime can indeed increase the dimensionality and help in increasing the performance of the PELM. For example, considering a regime modeled by $G_{sat}(x)=x/(I_{sat}+x)$ as suggested in previous works, e.g. \cite{pierangeli2021photonic}, can lead to the appearance of higher-order polynomial expansion of the input functions. Yet, two limitations may be noted in this case: first, for amplitude modulation, the activation function will still be of polynomial type, which is incompatible with the ELM requirements; second, by expanding $G_{sat}$, it is straightforward to conclude that higher-order corrections will become smaller and smaller, meaning that in real world conditions, may be obscured by experimental noise.

\subsection{Singular Value Decomposition: Methodology for Dimensionality Estimation and Noise Robustness}\label{sec:svd}

In practice, one way to quantify the dimensionality of the hidden space is to compute the numerical rank of $\boldsymbol{\bar{H}}$ via a singular value decomposition (SVD)~\cite{strang2019linear}
\begin{equation}
    \boldsymbol{\bar{H}} = \boldsymbol{\bar{U}} \boldsymbol{\bar{\Sigma}} \boldsymbol{\bar{V}}^T,
\label{eq: svd}
\end{equation}
where $\boldsymbol{\bar{U}}$ and $\boldsymbol{\bar{V}}$ are orthogonal matrices and $\boldsymbol{\bar{\Sigma}} = $ diag$(\sigma_1,\dots,\sigma_{N_{out}})$ contains the singular values ordered as $\sigma_1 \geq \dots \geq \sigma_{N_{out}}$. In theory, the number of non-zero singular values corresponds to the rank of the matrix $\boldsymbol{\bar{H}}$. Yet, in real-world conditions, experimental noise perturbs the data, introducing small singular values along directions that do not carry information.


To circumvent this limitation, we need to establish a threshold that defines a noise level for which the number of singular values exceeding it yields the estimated dimensionality. For this, we introduce the Weyl inequality that relates the SVD decomposition of the original matrix with one perturbed under additive noise $\boldsymbol{\bar{N}}$~\cite{horn1994topics} as
\begin{equation*}
    \| \sigma_i\left( \boldsymbol{\bar{H}} + \boldsymbol{\bar{N}}\right) - \sigma_i\left( \boldsymbol{\bar{H}} \right)\| \leq \sigma_1\left( \boldsymbol{\bar{N}}\right), \, i=1,\dots,N_{out}.
\end{equation*}
where $\sigma_1\left( \boldsymbol{\bar{N}}\right)$ is the largest singular value of the noise matrix. As for indices $i>rank\left(\boldsymbol{\bar{H}}\right)$ the true singular values vanish, we have that $\sigma_i\left( \boldsymbol{\bar{H}} + \boldsymbol{\bar{N}}\right) \leq \sigma_1\left( \boldsymbol{\bar{N}}\right)$, indicating that these modes lie below a noise threshold and can be discarded. This expression highlights $\sigma_1\left(\boldsymbol{\bar{N}}\right)$ as a viable threshold to identify singular values that belong to both true and noisy matrices. To estimate this threshold, we may then average several independent realisations of the dataset to compute the residual noise, and compute its SVD, see ~\ref{sec: rank}.
    
An additional feature of using the SVD decomposition is that it may also clarify the robustness of Ridge regression in the presence of noisy features. Indeed, substituting the decomposition of equation \ref{eq: svd} into the closed form Ridge solution in equation \ref{eq: ridge solution} and multiplying on the right by $\boldsymbol{\bar{H}}$ yields~\cite{hastie2009elements}
\begin{equation*}
    \boldsymbol{\beta}^* \boldsymbol{\bar{H}} = \boldsymbol{\bar{T}}\boldsymbol{\bar{V}} \left( \sum_{j=1}^{N_{out}} \frac{\sigma_j^2}{\sigma_j^2 + \lambda} \right) \boldsymbol{\bar{V}}_j^T
\label{eq: ridge svd effect}
\end{equation*}
where $\boldsymbol{\bar{V}}_j$ is the $j$-th column of $\boldsymbol{\bar{V}}$. The quantity in parentheses is often referred to as the effective degrees of freedom as the regularisation parameter $\lambda$ suppresses components with $\sigma_j^2 \ll \lambda$, effectively filtering noise, while directions with $\sigma_j^2 \gg \lambda$ are preserved with minimal distortion. Therefore, when optimising the $\lambda$ for best generalisation, we expect it to achieve values above the singular values induced by noise.
    

\section{\label{sec: experimental} Experimental Results and Model Validation}

Having set the theoretical model, we proceed to its validation using two proof-of-concept experiments on deliberately low-dimensional input spaces: a single-parameter regression task and a two-dimensional classification task involving non-linearly separable regions. Each experiment was executed both under amplitude and phase-only encoding to validate the lower bounds extracted from the model and demonstrate the limitations of the approach.

\subsection{Experimental Setup and Data Acquisition}

A simplified representation of the experimental setup is presented in Figure \ref{fig:OIComputer}. A $532nm$ diode laser with near-Gaussian wavefront is encoded by an SLM (Holoeye Pluto 2), enabling continuous feature encoding as described previously. A first polariser before the SLM ensures vertical polarisation, while a second polariser (an analyser) after the SLM enables either amplitude or phase modulation with a pre-calibrated linear response. The modulated beam is focused onto a diffusive medium with a $10\times$ microscope objective, generating a speckle pattern that is collected by a second $10\times$ objective and imaged onto a CMOS camera (Ximea MQ013MG-E2).

To investigate how the camera nonlinear response influences the hidden space dimensionality, we vary the sensor exposure time so that the detector operates either in a nearly linear regime or in saturation. For amplitude-only encoding, we set exposure times to $2ms$ and $30ms$, whereas for phase-only encoding, we use $0.5ms$ and $7ms$. This controlled variation allows us to isolate the effect of sensor nonlinearity on the effective dimensionality of the hidden layer. In total, for each experiment - regression or classification task - there are four datasets with the settings amplitude or phase encoding with low or high exposure.

The hidden layer output matrix dataset is obtained by encoding one input $\boldsymbol{X}_i$ at a time, propagating it through the system, and subsequently recording the associated speckle at the camera plane corresponding to one row of $\boldsymbol{\bar{H}}$. To characterise the hidden space dimensionality in the presence of environmental noise according to the method described in section \ref{sec:svd}, we record ten independent datasets for each configuration. This approach serves two purposes. First, it allows for the estimation of the Weyl threshold for identifying relevant singular values. Second, training on one dataset and evaluating with another allows the analysis of the robustness of the model to experimental noise. 
    
Model evaluation is performed in the following way. The ten independent datasets acquired are distributed into one for training the model, four for validation, and five for testing. Within each dataset, we perform a 5-fold cross-validation, corresponding to an $80-20\%$ split of feature points into train-validation subsets. The train, validation, and test root-mean-square error (RMSE) and accuracy metrics presented are the results of the average over the 5-folds and respective dataset distribution. Ridge regression is implemented with scikit-learn, and its regularisation hyperparameter $\lambda$ is tuned on the four validation sets by minimising RMSE (for regression) or maximising accuracy (for classifications). The final test scores are obtained from the validation points of the datasets withheld from both training and validation, preventing any data leakage to bias the results.

\subsection*{Performance in Regression Task}

To evaluate the system on a simple regression task, we try to approximate the nonlinear function $g\left(x\right) = \mathrm{sinc}\left(x\right)$ in a one-dimensional input space ($N_{I}=1$), corresponding to $N_{enc}=2$ spatial modes. According to equations \ref{eq: amp rank} and \ref{eq: phs rank}, this setting yields a theoretical hidden space dimensionality of $rank\left(\boldsymbol{\bar{H}}\right) = 3$ for both amplitude-only and phase-only encoding schemes. The input domain $x\in\left[-8,8\right]$ is normalised to match the SLM modulation range and discretised into 64 equally spaced values.
    
\begin{figure}
    \centering
    \makebox[\textwidth]{
    \includegraphics[width=1.\textwidth]{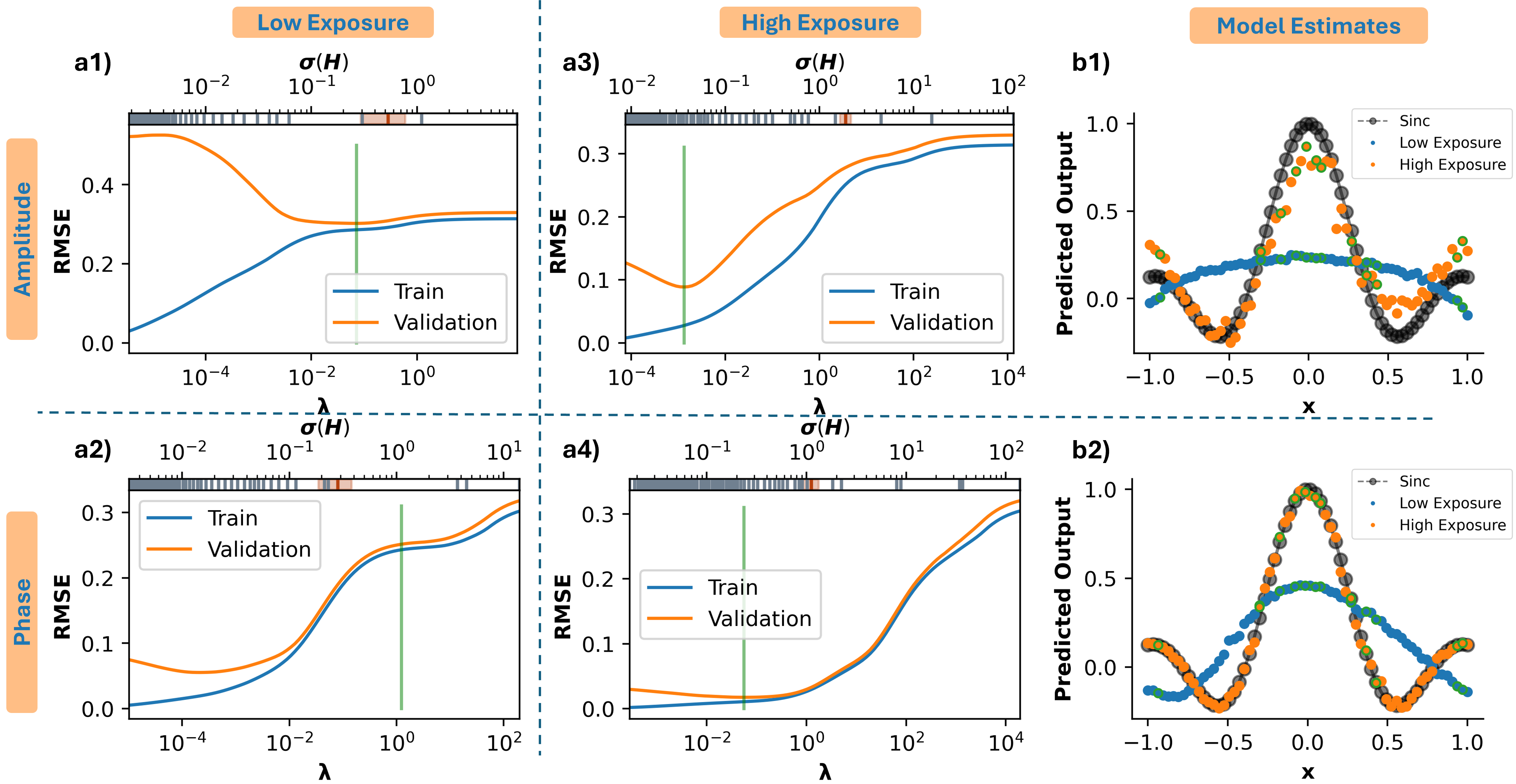}
    }
    \caption{Regression results. (a) Train and validation RMSE versus ridge parameter $\lambda$ for amplitude (top row) and phase (bottom row) encoding under low (left) and high (right) exposure settings. Gray bars above each plot show the singular value spectrum of the hidden layer such that $\sigma^2 = \lambda$, the red band marks the Weyl noise threshold. The green marker denotes the $\lambda$ value chosen for testing. (b) Single fold predictions for the low and high exposure settings compared with the ground truth sinc function (dashed black) for amplitude (top row) and phase (bottom row) modulations. A green edge denotes the test points while the training points have no edge. For both modulation settings, the low exposure configuration showcases the limited expressivity of the model to the underlying encoding functions while the high exposure nonlinearity enriches the effective hidden space, resulting in higher expressivity, as reflected in an improved fit between prediction and target function.}
    \label{fig: regression results}
\end{figure}

In Figure~\ref{fig: regression results}.a1, for amplitude encoding under low exposure (first row and column), the camera response is approximately linear, resulting in three dominant singular components above the minimum value of the Weyl threshold (red region) as expected. For improved observation of the single value distribution and Weyl threshold we plotted their values in the ~\ref{sec: rank}, where it can be seen that the third largest SVD is already in the bottom region of the estimated threshold. These suggest that not only the approximation capabilities may be affected (due to low dimensionality of the hidden space) but that noise may also strongly affect performance. Indeed, in top panel of Figure~\ref{fig: regression results}.b1 it is easy to observe that the predictions of the PELM in low exposure regime (blue markers) cannot approximate anything beyond second-order polynomials, as expected by their activation functions. 

A similar behavior is obtained for phase-only encoding with low exposure, Figure~\ref{fig: regression results}.a2, with the empirical Weyl noise level estimate suggesting now a five-dimensional hidden-space rank instead of the expected three. This may indicate that some nonlinearity may be introduced by residual sensor nonlinearity caused by higher intensities throughout the variation of the input phase, or even due to the deadband of the camera in lower intensity regions. However, we shall also note that the gap between the fourth and third singular value is of one order of magnitude, reflecting the much higher participation of the expected three components. Despite the presence of sensor-induced higher-order contributions, we can still tune $\lambda$ to filter the smaller contributions and still study the system in an approximation to the linear regime. For this, we set the hyperparameter $\lambda \approx 4$ penalising the smaller components, to obtain the results present in Figure \ref{fig: regression results} b2). Again, it is easy to see that the predictions cannot approximate the target function, which is again connected to the lack of dimensionality of the output space.


Finally, to increase the dimensionality of the output space what we can do is to increase the camera exposure time, entering a nonlinear saturable regime. In Figure~\ref{fig: regression results}.a3, for amplitude, the optimal $\lambda$ now includes a higher number of singular components, while for phase in Figure~\ref{fig: regression results}.a4, this can also be seen in the growth of singular components identified by the Weyls threshold to $8$. Quantitatively, this lowers the test RMSE from $0.31$ for low exposure to $0.13$ for amplitude encoding, and from $0.25$ to $0.02$ for phase-only modulation. Qualitatively, in Figure~\ref{fig: regression results}.b1 for amplitude and Figure~\ref{fig: regression results}.b2 for phase modulation, the models expressivity is seen to allow improved fitting of predictions to the target function, thus leaving their polynomial-like limitation.
    
\subsection*{Performance in Classification Tasks}

The classification experiment explores a two-dimensional point distribution along two distinct spirals, resulting in an encoding space with $N_{enc}=3$. According to equations \ref{eq: amp rank} and \ref{eq: phs rank}, we expect a hidden space dimensionality of $6$ for amplitude-only encoding and $7$ for phase-only encoding. Input coordinates $x, y \in \left[-3,3\right]$ are generated with fixed noise along two spiral functions with 32 sampled points for each spiral.

\begin{figure}
    \makebox[\textwidth]{
    \includegraphics[width=1.\textwidth]{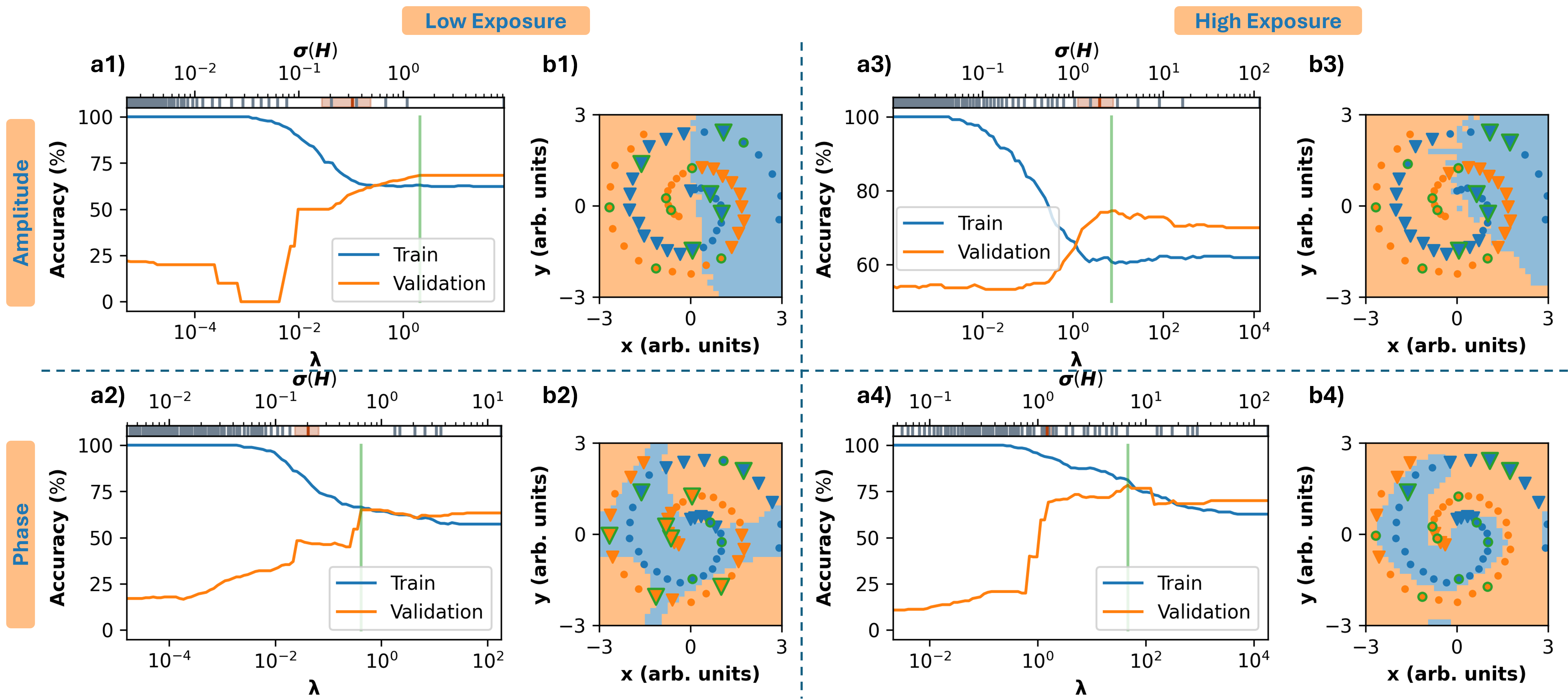}
    }
    \caption{Classification results. Low and high exposure versus amplitude and phase encoding table with each entry showing an accuracy versus $\lambda$ plot. The green marker denotes the optimal $\lambda$ used for evaluation. Alongside each plot is an example of the predictions of the train (no edges) and test (green edge) data points from a single fold. Correct predictions are marked with a dot, while incorrect predictions have a triangular shape. The background of these figures is the prediction map over a $32\times 32$ grid showcasing the prediction field used to analyse the generalisation capabilities of the model.}
\label{fig: classification results}
\end{figure}

Under amplitude encoding and low exposure, the singular value distribution and Weyl-based noise level estimate are shown in Figure~\ref{fig: classification results}.a indicating the presence of $5$  singular components, one below the theoretical prediction. A corresponding jump in classification accuracy occurs when $\lambda$ is chosen between the fifth and sixth singular values, corroborating an effective rank of five. The missing component remains unobserved due to the lack of variability in the mixed term $xy$, i.e. when either coordinate is small, the signal may be low and indistinguishable from noise. As a consequence of the limited dimensionality and expressivity, the test and validation accuracies for the optimal $\lambda$ are $67\%$ and $68\%$, respectively, and the model cannot generalize properly.

Under phase-only encoding and low exposure, the empirical Weyl noise level is shown in Figure~\ref{fig: classification results}.a2 (bottom row, see~\ref{sec: rank} for a clearer representation) and the cross-validation correctly estimate the expected seven significant singular components. Unlike amplitude-only encoding, phase modulation redirects the light rather than controlling the intensity, rendering the speckle intensity approximately invariant to input features, resulting in improved observation of mixed terms. Nevertheless, despite this higher hidden-space dimensionality, due to limited expressivity, the optimal test and validation accuracies remain only at $65\%$, similar to the amplitude-only case.
    
Finally, when increasing the camera exposure to induce stronger nonlinearities, we observe improved classification performance in both encoding schemes. In the case of amplitude encoding, the rank grows from $5$ in low exposure to $7$ for high exposure, while for phase, it grows from $7$ to $22$. The cross-validation optimal model yields test and validation accuracies of $75\%$ for amplitude-only encoding and $78\%$ for phase-only encoding. The generalisation map in Figure~\ref{fig: classification results}.b4 reveal the key difference. Under amplitude-only encoding in Figure~\ref{fig: classification results}.b1 and~\ref{fig: classification results}.b3 or phase encoding with low exposure in Figure~\ref{fig: classification results}.b2, predictions fail to capture the spiral geometry. In contrast, phase-only predictions with high exposure are shown in Figure~\ref{fig: classification results}.b4 begin to conform to the spiral distribution, indicating a more accurate representation of the underlying structure.
    
\section{\label{sec: discussion} Discussion and Concluding Remarks}


Overall, the model and results presented provide a quantitative link between coherent light propagation and the theory of ELMs. By describing each stage of PELM separately, input encoding, complex-valued linear propagation, and detection, we can isolate the contribution of each block and how they interplay. Casting the free-space optical processor as an ELM with a complex-valued linear layer places this architecture in the family of C-ELMs, bringing the analytical tools of that framework to photonic hardware.

However, the model shows that the function space of the hidden space is ultimately limited by the input encoding functions. With a purely linear camera response, the network can only form polynomials of the pairwise combinations of input encoding functions. As a result, the hidden layer dimensionality is upper-bounded and depends solely on the number of input modes, failing to show the expected universal approximation capabilities predicted for ELM. For amplitude and phase modulation, taken together or separately, we derived closed-form expressions for that bound, capturing the underlying combinatorial process. Operating the experiment at low exposures confirmed these predictions, where we estimate the dimensionality of the system using a Weyl estimation of noise level, used to exclude residual components. Increasing the detector exposure, in turn, introduced additional non‑linearity, enriching the hidden space and lifting several more singular values above the noise floor. 
    
These results imply that enriching the input space of functions, through pre-processing or multiple encodings, cannot overcome the dimensionality limitation and therefore falls short of the universal approximation property. Modifying the detector function, here achieved by raising the exposure time, does increase the effective dimensionality, pointing to a practical route towards a more expressive PELM framework. However, the experimental observation of the full-rank condition for saturable nonlinearity on the detection side was not possible due to the presence of noise.

In summary, this manuscript clarifies the relationship between the standard ELM algorithm and its photonic implementation. While PELMs have shown limitations, they remain attractive thanks to their versatility and energy efficiency for tasks that profit from rapid, hardware-level feature expansion, in particular if one adds a nonlinear propagation layer instead of a simple diffusive layer\cite{silva2024optical,azam2024optically}.

\section*{Acknowledgements}

This work is co-financed by Component 5 - Capitalization and Business Innovation, integrated in the Resilience Dimension of the Recovery and Resilience Plan within the scope of the Recovery and Resilience Mechanism (MRR) of the European Union (EU), framed in the Next Generation EU, for the period 2021 - 2026, within project HfPT, with
reference 41, and by national funds through FCT – Fundação para a Ciência e a Tecnologia, I.P., under the support UID/50014/2023 (https://doi.org/10.54499/UID/50014/2023). Nuno A. Silva acknowledges the support of FCT under the grant 2022.08078.CEECIND/CP1740/CT0002 (https://doi.org/10.54499/2022.08078.CEECIND/CP1740/CT0002). 

\section*{Author contributions}

V.R. theoretical model, analysed results and wrote manuscript; N.A.S. conceived experiments, theoretical model, and wrote manuscript; D.S. theoretical model; F.C.M. conducted the experiments, analysed results and reviewed the manuscript; T.D.F. conducted the experiments and reviewed the manuscript; C.S.M. reviewed the manuscript.

\section*{Data Availability}
Data underlying the results presented in this paper are not publicly available at this time but may be obtained from the authors upon reasonable request.

\bibliographystyle{unsrt}
\bibliography{mybib}

\begin{thebibliography}{10}

\bibitem{athale2016optical}
Ravi Athale and Demetri Psaltis.
\newblock Optical computing: past and future.
\newblock {\em Optics and Photonics News}, 27(6):32--39, 2016.

\bibitem{richardson2013space}
David~J Richardson, John~M Fini, and Lynn~E Nelson.
\newblock Space-division multiplexing in optical fibres.
\newblock {\em Nature photonics}, 7(5):354--362, 2013.

\bibitem{goodman2005introduction}
Joseph~W Goodman.
\newblock {\em Introduction to Fourier optics}.
\newblock Roberts and Company publishers, 2005.

\bibitem{lugt1964signal}
A~Vander Lugt.
\newblock Signal detection by complex spatial filtering.
\newblock {\em IEEE Transactions on information theory}, 10(2):139--145, 1964.

\bibitem{athale1982optical}
RA~Athale, HH~Szu, and JN~Lee.
\newblock Optical implementation of integral transforms with bessel function kernels.
\newblock {\em Optics Letters}, 7(3):124--126, 1982.

\bibitem{shen2017deep}
Yichen Shen, Nicholas~C Harris, Scott Skirlo, Mihika Prabhu, Tom Baehr-Jones, Michael Hochberg, Xin Sun, Shijie Zhao, Hugo Larochelle, Dirk Englund, et~al.
\newblock Deep learning with coherent nanophotonic circuits.
\newblock {\em Nature photonics}, 11(7):441--446, 2017.

\bibitem{spall2020fully}
James Spall, Xianxin Guo, Thomas~D Barrett, and AI~Lvovsky.
\newblock Fully reconfigurable coherent optical vector--matrix multiplication.
\newblock {\em Optics Letters}, 45(20):5752--5755, 2020.

\bibitem{matthes2019optical}
Maxime~W Matth{\`e}s, Philipp Del~Hougne, Julien De~Rosny, Geoffroy Lerosey, and S{\'e}bastien~M Popoff.
\newblock Optical complex media as universal reconfigurable linear operators.
\newblock {\em Optica}, 6(4):465--472, 2019.

\bibitem{chang2018hybrid}
Julie Chang, Vincent Sitzmann, Xiong Dun, Wolfgang Heidrich, and Gordon Wetzstein.
\newblock Hybrid optical-electronic convolutional neural networks with optimized diffractive optics for image classification.
\newblock {\em Scientific reports}, 8(1):12324, 2018.

\bibitem{wu2021programmable}
Changming Wu, Heshan Yu, Seokhyeong Lee, Ruoming Peng, Ichiro Takeuchi, and Mo~Li.
\newblock Programmable phase-change metasurfaces on waveguides for multimode photonic convolutional neural network.
\newblock {\em Nature communications}, 12(1):96, 2021.

\bibitem{kazanskiy2022optical}
Nikolay~L Kazanskiy, Muhammad~A Butt, and Svetlana~N Khonina.
\newblock Optical computing: Status and perspectives.
\newblock {\em Nanomaterials}, 12(13):2171, 2022.

\bibitem{hopfield1982neural}
John~J Hopfield.
\newblock Neural networks and physical systems with emergent collective computational abilities.
\newblock {\em Proceedings of the national academy of sciences}, 79(8):2554--2558, 1982.

\bibitem{farhat1985optical}
Nabil~H Farhat, Demetri Psaltis, Aluizio Prata, and Eung Paek.
\newblock Optical implementation of the hopfield model.
\newblock {\em Applied optics}, 24(10):1469--1475, 1985.

\bibitem{hornik1991approximation}
Kurt Hornik.
\newblock Approximation capabilities of multilayer feedforward networks.
\newblock {\em Neural networks}, 4(2):251--257, 1991.

\bibitem{wetzstein2020inference}
Gordon Wetzstein, Aydogan Ozcan, Sylvain Gigan, Shanhui Fan, Dirk Englund, Marin Solja{\v{c}}i{\'c}, Cornelia Denz, David~AB Miller, and Demetri Psaltis.
\newblock Inference in artificial intelligence with deep optics and photonics.
\newblock {\em Nature}, 588(7836):39--47, 2020.

\bibitem{lin2018all}
Xing Lin, Yair Rivenson, Nezih~T Yardimci, Muhammed Veli, Yi~Luo, Mona Jarrahi, and Aydogan Ozcan.
\newblock All-optical machine learning using diffractive deep neural networks.
\newblock {\em Science}, 361(6406):1004--1008, 2018.

\bibitem{luo2022metasurface}
Xuhao Luo, Yueqiang Hu, Xiangnian Ou, Xin Li, Jiajie Lai, Na~Liu, Xinbin Cheng, Anlian Pan, and Huigao Duan.
\newblock Metasurface-enabled on-chip multiplexed diffractive neural networks in the visible.
\newblock {\em Light: Science \& Applications}, 11(1):158, 2022.

\bibitem{liu2022programmable}
Che Liu, Qian Ma, Zhang~Jie Luo, Qiao~Ru Hong, Qiang Xiao, Hao~Chi Zhang, Long Miao, Wen~Ming Yu, Qiang Cheng, Lianlin Li, et~al.
\newblock A programmable diffractive deep neural network based on a digital-coding metasurface array.
\newblock {\em Nature Electronics}, 5(2):113--122, 2022.

\bibitem{zheng2022optimize}
Minjia Zheng, Lei Shi, and Jian Zi.
\newblock Optimize performance of a diffractive neural network by controlling the fresnel number.
\newblock {\em Photonics Research}, 10(11):2667--2676, 2022.

\bibitem{abou2025programmable}
Loubnan Abou-Hamdan, Emil Marinov, Peter Wiecha, Philipp del Hougne, Tianyu Wang, and Patrice Genevet.
\newblock Programmable metasurfaces for future photonic artificial intelligence.
\newblock {\em Nature Reviews Physics}, pages 1--17, 2025.

\bibitem{huang2004extreme}
Guang-Bin Huang, Qin-Yu Zhu, and Chee-Kheong Siew.
\newblock Extreme learning machine: a new learning scheme of feedforward neural networks.
\newblock In {\em 2004 IEEE international joint conference on neural networks (IEEE Cat. No. 04CH37541)}, volume~2, pages 985--990. Ieee, 2004.

\bibitem{Huang2006ExtremeLM}
Guangbin Huang, Qin-Yu Zhu, and Chee~Kheong Siew.
\newblock Extreme learning machine: Theory and applications.
\newblock {\em Neurocomputing}, 70:489--501, 2006.

\bibitem{li2005fully}
Ming-Bin Li, Guang-Bin Huang, Paramasivan Saratchandran, and Narasimhan Sundararajan.
\newblock Fully complex extreme learning machine.
\newblock {\em Neurocomputing}, 68:306--314, 2005.

\bibitem{pierangeli2021photonic}
Davide Pierangeli, Giulia Marcucci, and Claudio Conti.
\newblock Photonic extreme learning machine by free-space optical propagation.
\newblock {\em Photonics Research}, 9(8):1446--1454, 2021.

\bibitem{m2022large}
Carlo M.~Valensise, Ivana Grecco, Davide Pierangeli, and Claudio Conti.
\newblock Large-scale photonic natural language processing.
\newblock {\em Photonics Research}, 10(12):2846--2853, 2022.

\bibitem{redding2024fiber}
Brandon Redding, Joseph~B Murray, Joseph~D Hart, Zheyuan Zhu, Shuo~S Pang, and Raktim Sarma.
\newblock Fiber optic computing using distributed feedback.
\newblock {\em Communications Physics}, 7(1):75, 2024.

\bibitem{biasi2023array}
Stefano Biasi, Riccardo Franchi, Lorenzo Cerini, and Lorenzo Pavesi.
\newblock An array of microresonators as a photonic extreme learning machine.
\newblock {\em APL Photonics}, 8(9), 2023.

\bibitem{rausell2025programmable}
Jos{\'e}~Roberto Rausell-Campo, Antonio Hurtado, Daniel P{\'e}rez-L{\'o}pez, and Jos{\'e} Capmany~Francoy.
\newblock Programmable photonic extreme learning machines.
\newblock {\em Laser \& Photonics Reviews}, 19(9):2400870, 2025.

\bibitem{liutkus2014imaging}
Antoine Liutkus, David Martina, S{\'e}bastien Popoff, Gilles Chardon, Ori Katz, Geoffroy Lerosey, Sylvain Gigan, Laurent Daudet, and Igor Carron.
\newblock Imaging with nature: Compressive imaging using a multiply scattering medium.
\newblock {\em Scientific reports}, 4(1):5552, 2014.

\bibitem{saade2016random}
Alaa Saade, Francesco Caltagirone, Igor Carron, Laurent Daudet, Ang{\'e}lique Dr{\'e}meau, Sylvain Gigan, and Florent Krzakala.
\newblock Random projections through multiple optical scattering: Approximating kernels at the speed of light.
\newblock In {\em 2016 IEEE International Conference on Acoustics, Speech and Signal Processing (ICASSP)}, pages 6215--6219. IEEE, 2016.

\bibitem{popoff2010measuring}
S{\'e}bastien~M Popoff, Geoffroy Lerosey, R{\'e}mi Carminati, Mathias Fink, Albert~Claude Boccara, and Sylvain Gigan.
\newblock Measuring the transmission matrix in optics: An approach to the study and control<? format?> of light propagation in disordered media.
\newblock {\em Physical review letters}, 104(10):100601, 2010.

\bibitem{hastie2009elements}
Trevor Hastie, Robert Tibshirani, Jerome Friedman, et~al.
\newblock The elements of statistical learning, 2009.

\bibitem{bartlett2002sample}
Peter~L Bartlett.
\newblock The sample complexity of pattern classification with neural networks: the size of the weights is more important than the size of the network.
\newblock {\em IEEE transactions on Information Theory}, 44(2):525--536, 2002.

\bibitem{strang2019linear}
Gilbert Strang et~al.
\newblock {\em Linear algebra and learning from data}, volume~4.
\newblock Wellesley-Cambridge Press Cambridge, 2019.

\bibitem{horn1994topics}
Roger~A Horn and Charles~R Johnson.
\newblock {\em Topics in matrix analysis}.
\newblock Cambridge university press, 1994.

\bibitem{silva2024optical}
Nuno~A Silva, Vicente Rocha, and Tiago~D Ferreira.
\newblock Optical extreme learning machines with atomic vapors.
\newblock {\em Atoms}, 12(2):10, 2024.

\bibitem{azam2024optically}
Pierre Azam and Robin Kaiser.
\newblock Optically accelerated extreme learning machine using hot atomic vapors.
\newblock {\em Physical Review Applied}, 22(3):034041, 2024.

\end{thebibliography}

\appendix

\section{\label{ELM framework} ELM framework}

    The ELM algorithm is summarised in three steps~\cite{huang2004extreme}, described as follows. First, it assigns an arbitrarily distributed weight matrix $\boldsymbol{\bar{w}}$ and bias vector $\boldsymbol{b}$ which can be real or complex valued in the case of complex ELMs~\cite{li2005fully}, producing a random, nonlinear mapping of the input features $\boldsymbol{X} = \left[x_1, \dots, x_N\right]^T$ into the hidden space via the transformation $g\left(\boldsymbol{\bar{w}}\boldsymbol{X} + \boldsymbol{b}\right)$ where $g$ is the element-wise nonlinear activation function. Second, it constructs the hidden layer output matrix with $N_D$ training samples along the rows and $N$ hidden nodes along the columns,
\begin{equation}
    \boldsymbol{\bar{\mathcal{H}}} = \left[ \begin{array}{ccc}
    g\left(\boldsymbol{\bar{w}}_1\boldsymbol{X_1} + \boldsymbol{b}\right) & \dots & g\left(\boldsymbol{\bar{w}}_N\boldsymbol{X_1} + \boldsymbol{b}\right)  \\
    \vdots & \dots & \vdots \\
    g\left(\boldsymbol{\bar{w}}_1\boldsymbol{X_{N_D}} + \boldsymbol{b}\right) & \dots & g\left(\boldsymbol{\bar{w}}_N\boldsymbol{X_{N_D}} + \boldsymbol{b}\right)
\end{array}    
\right]
\label{output_matrix}
\end{equation}
    where the index of $\boldsymbol{\bar{w}}$ is the row of the weight matrix. In the final step, the hidden layer output matrix is used to compute the linear readout $\boldsymbol{\beta}$ solving $\boldsymbol{\bar{\mathcal{H}}}\boldsymbol{\beta} = \boldsymbol{T}$, where $\boldsymbol{T}$ contains the training labels. This makes the hidden layer output matrix the core subject in the universal approximation proof for ELMs~\cite{Huang2006ExtremeLM}. By guaranteeing that it has full rank independently of the network size, the inverse of $\mathcal{\boldsymbol{\bar{H}}}$ always exists, guaranteeing the linear regression has a solution. The proof of this theorem leverages the randomness of the weights, such that $\boldsymbol{\bar{w_j}} \boldsymbol{X}_k \neq \boldsymbol{\bar{w_j}} \boldsymbol{X}_k'$ for all $k \neq k'$ and the infinitely differentiable activation function resulting in a proof where the columns of the matrix cannot belong to any space with fewer dimensions than N.

\section{\label{sec: rank} Rank Distributions}

    To estimate the singular value spectrum and the Weyls threshold we adopt the following procedure. Let $\mathcal{D}=\{\bar{\mathbf{H}}^{(i)}\}_{i=1}^{M}$ be the set of hidden layer output matrices obtained from $M$ independent experimental realisations. For each realisation we compute the residual by subtracting the ensemble mean,
\begin{equation*}
    \boldsymbol{\bar{N}}^i = \boldsymbol{\bar{H}}^i - \frac{1}{N}\sum_{j=1}^M \boldsymbol{\bar{H}}^j,
\end{equation*}
    where $i=1,\dots, N$. A SVD of each $\boldsymbol{\bar{N}}^i$ and $\boldsymbol{\bar{H}}^{i}$ yield the largest singular value of the residual $\sigma_{1}^{i}\left(\boldsymbol{\bar{N}}\right)$ and the spectra of singular values $\sigma_j^i\left(\boldsymbol{\bar{H^i}}\right)$ for $j=1,\dots, min\left(N_{out}, N_{D}\right)$, respectively. By averaging $\sigma_{1}^{i}\left(\boldsymbol{\bar{N}}\right)$ and $\sigma_j^i\left(\boldsymbol{\bar{H^i}}\right)$ for $j=1,\dots, min\left(N_{out}, N_{D}\right)$ over the $M$ realisations we obtain the Weyl threshold and spectra used for analysis.

\begin{figure}
    \centering
    \includegraphics[width=1.\linewidth]{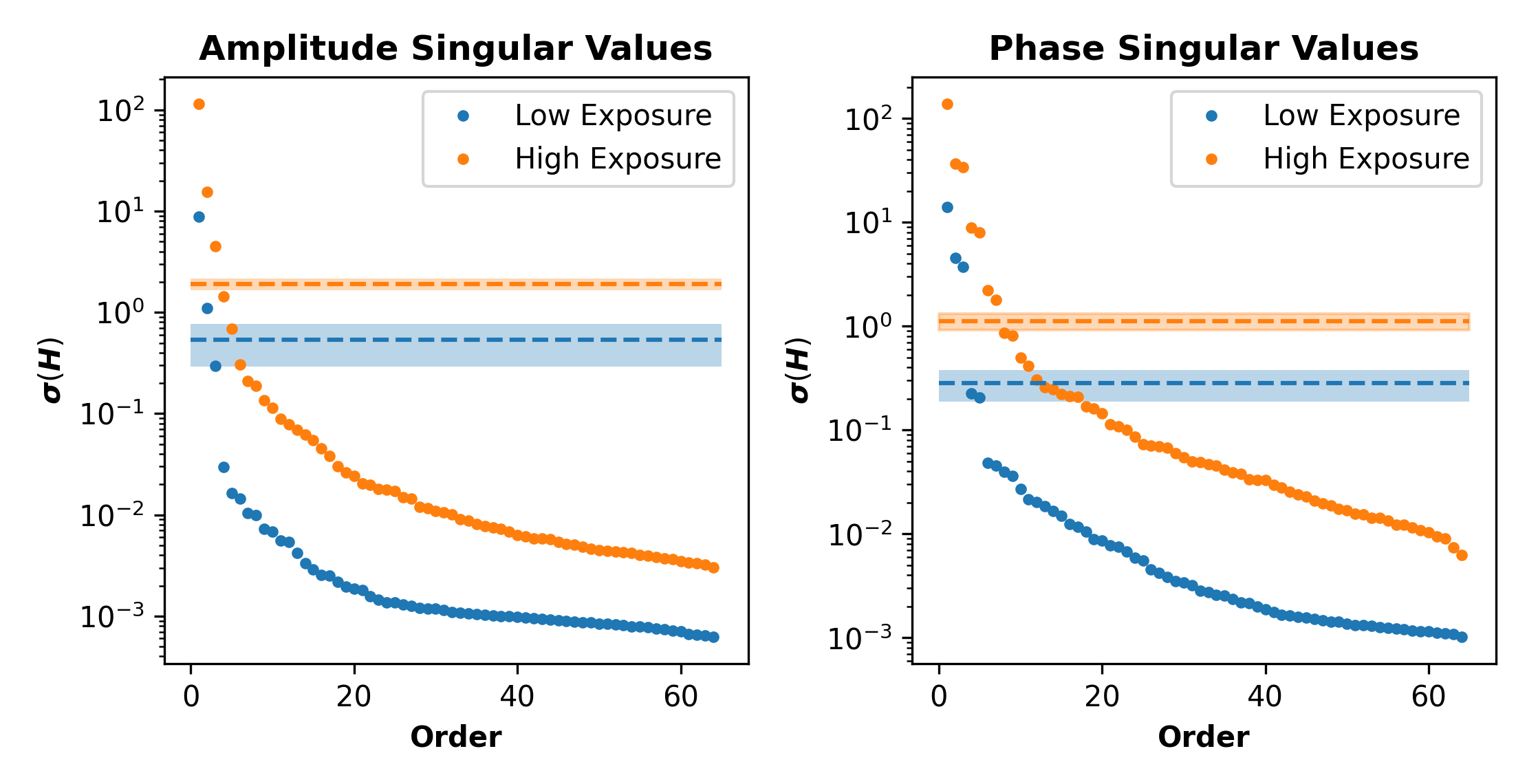}
    \caption{Singular value spectra and Weyl threshold for the regression configuration. In the left amplitude encoding spectra for low and high exposure, identifying $3$ singular values above the Weyl threshold (highlighted region). Similar setup on the right for phase encoding we identify $5$ and $8$ singular values above the threshold.}
    \label{fig: rank regression}
\end{figure}

    The results for the regression task in Figure~\ref{fig: rank regression} are obtained with this procedure. For amplitude encoding with low or high exposure we denote $3$ singular values above the Weyl threshold while for phase we identify $5$ and $8$, respectively.
    
\begin{figure}
    \centering
    \includegraphics[width=1.\linewidth]{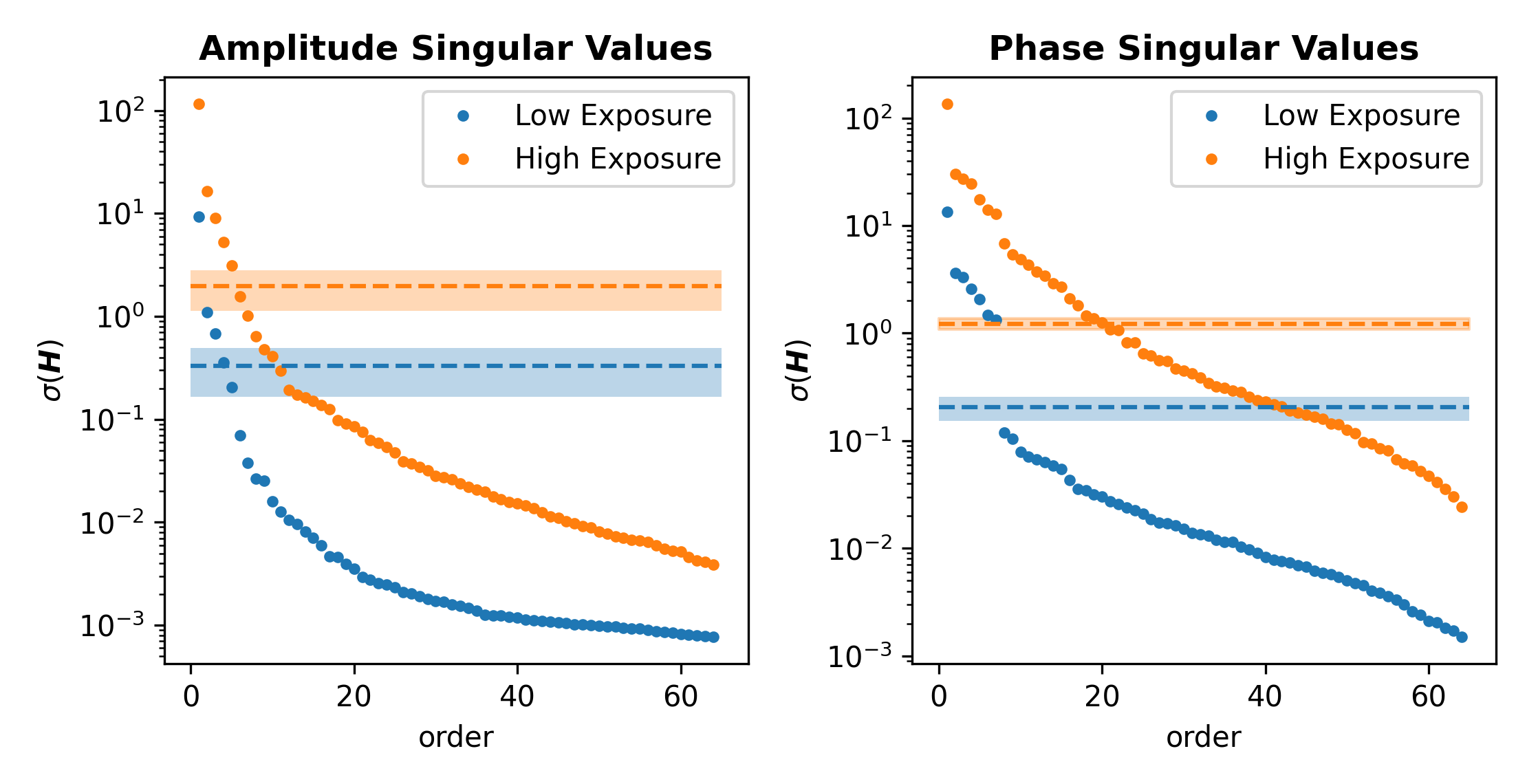}
    \caption{Singular value spectra and Weyl threshold for the classification configuration. In the left amplitude encoding spectra for low and high exposure, identifying $5$ and $6$ singular values above the Weyl threshold (highlighted region). Similar setup on the right for phase encoding we identify $7$ and $22$ singular values above the threshold.}
    \label{fig: rank classification}
\end{figure}

    In the classification task with results presented in Figure~\ref{fig: rank classification}, low exposure yields $5$ and $7$ significant singular values for amplitude and phase encoding, while high exposure increases these numbers to $6$ and $22$, respectively.

\end{document}